\documentclass{webofc}
\usepackage[varg]{txfonts}   
\usepackage{subfigure} 
\newcommand{\eps}{\epsilon}

\newcommand{\beq}{\begin{equation}}
\newcommand{\eeq}{\end{equation}}
\newcommand{\ba}{\begin{array}}
\newcommand{\ea}{\end{array}}
\newcommand{\bea}{\begin{eqnarray}}
\newcommand{\eea}{\end{eqnarray}}
\newcommand{\bi}{\begin{itemize}}  
\newcommand{\ei}{\end{itemize}}
\newcommand{\ben}{\begin{enumerate}} 
\newcommand{\een}{\end{enumerate}}
\newcommand{\bc}{\begin{center}}
\newcommand{\ec}{\end{center}}

\newcommand{\txt}{\textstyle}

\newcommand{\third}{{\txt \frac{1}{3}}}

\newcommand{\Tr}{{\rm Tr}}
\newcommand{\Lagr}{\mathcal{L}}

\usepackage[normalem]{ulem}  

\begin{document}
\title{Dynamics of semi-superfluid fluxtubes in color-flavor locked quark matter}
\woctitle{CONF12}
%
%

\author{\firstname{Mark G.} \lastname{Alford}\inst{1}\fnsep\thanks{\email{alford@physics.wustl.edu}} \and
        \firstname{Andreas} \lastname{Windisch}\inst{1}\fnsep\thanks{\email{windisch@physics.wustl.edu}}
}

\institute{Physics Department, Washington University, St. Louis, MO 63130, USA}

\abstract{
At very high densities, as for example in the core of a neutron star, matter may appear in the color-flavor locked (CFL) phase, which is a superfluid. This phase features topologically stable vortex solutions, which arise in a spinning superfluid as localized configurations carrying quanta of angular momentum. 
 Despite the topological stability of these vortices they are not the lowest energy state of the system at neutron star densities and decay into triplets of semi-superfluid fluxtubes. In these proceedings we report on the progress of our numerical study in the Ginzburg-Landau approach, where we investigate lattices of semi-superfluid fluxtubes.  The fluxtubes are obtained through controlled decay of global vortex configurations in the presence of a gauge field. Understanding the dynamics of semi-superfluid string configurations is important in the context of angular momentum transfer from a quark matter core of a neutron star beyond the core boundary, since not vortex-, but fluxtube pinning seems to be the relevant mechanism in this scenario.
}
\maketitle
%
\section{Introduction}
\label{intro}
\begin{figure*}[ht!]
\centering
\subfigure[][]{
 \label{fig:decay_a}
\includegraphics[width=0.42\hsize]{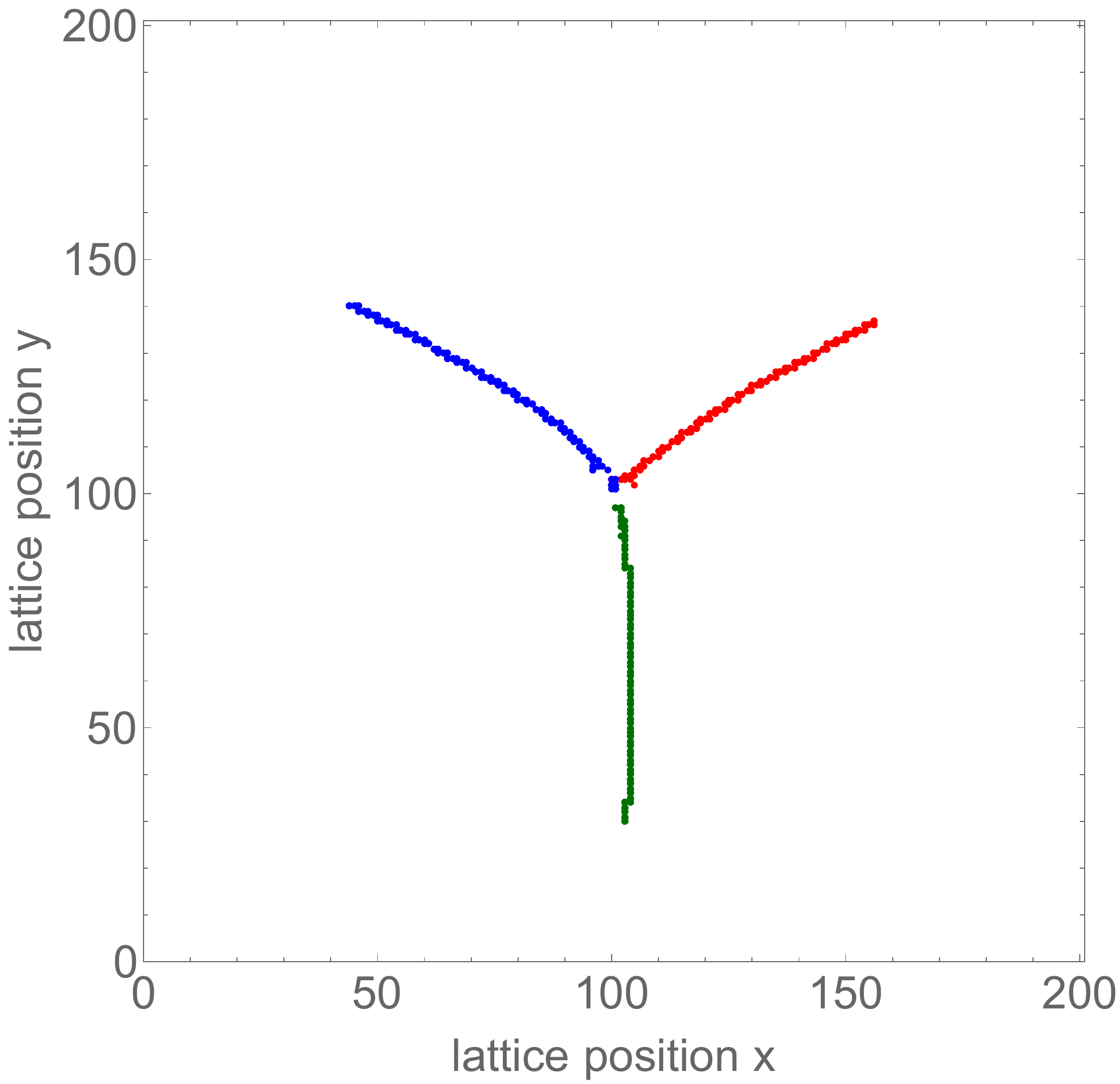}
}\hspace{8pt}
\subfigure[][]{
 \label{fig:decay_b}
\includegraphics[width=0.40\hsize]{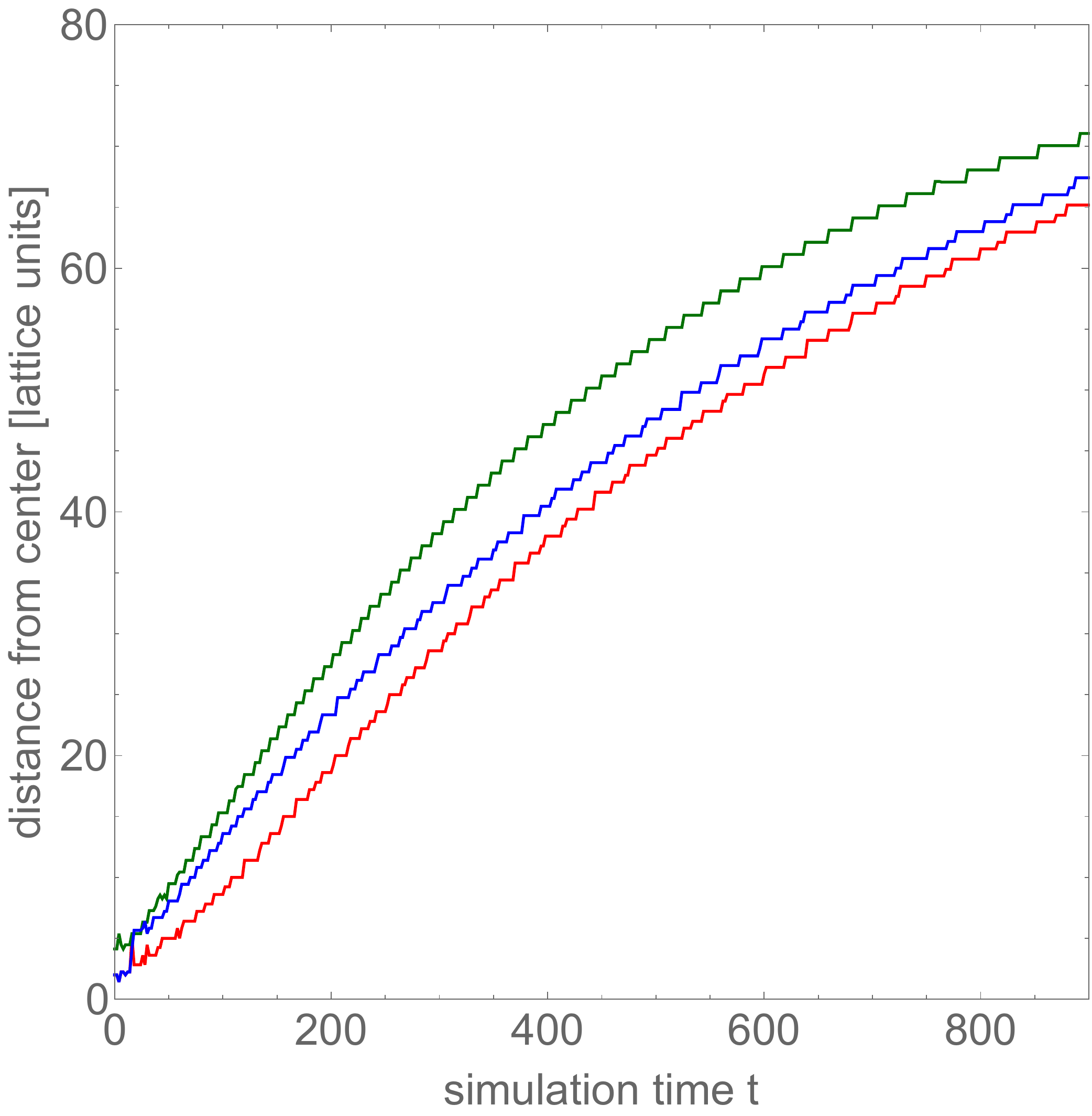}
}
\caption[]{Typical decay of a global vortex configuration into three semi-superfluid fluxtubes. Panel \ref{fig:decay_a} shows the paths of the semi-superfluid fluxtubes as they travel away from one another. The fluxtubes are the decay products of a vortex that was placed at the center of the lattice. The travel paths are slightly bent, which is due to the fact that the decay did not occur in a very symmetric manner. Panel \ref{fig:decay_b} shows the distance of each vortex from the center as a function of time. The non-smooth behavior is due to the discretization of the lattice, and the oscillatory behavior at $t=0$ due to the (arbitrarily chosen) criterion of decay detection in the simulation code.}
\label{fig:decay}
\end{figure*}

\begin{figure*}[ht!]
\centering
\subfigure[][]{
 \label{fig:decay_contour_a}
\includegraphics[width=0.30\hsize]{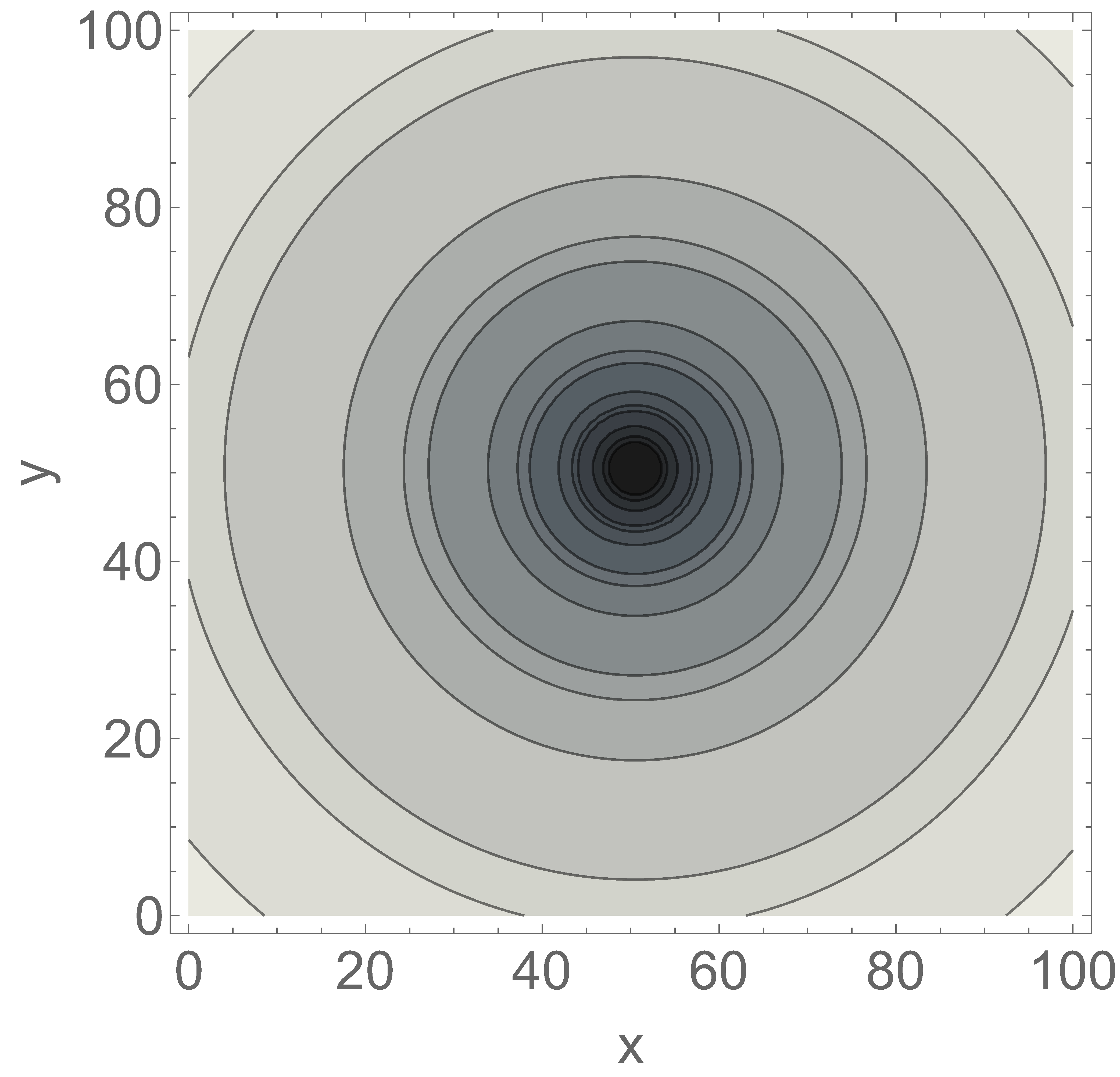}
}\hspace{8pt}
\subfigure[][]{
 \label{fig:decay_contour_b}
\includegraphics[width=0.30\hsize]{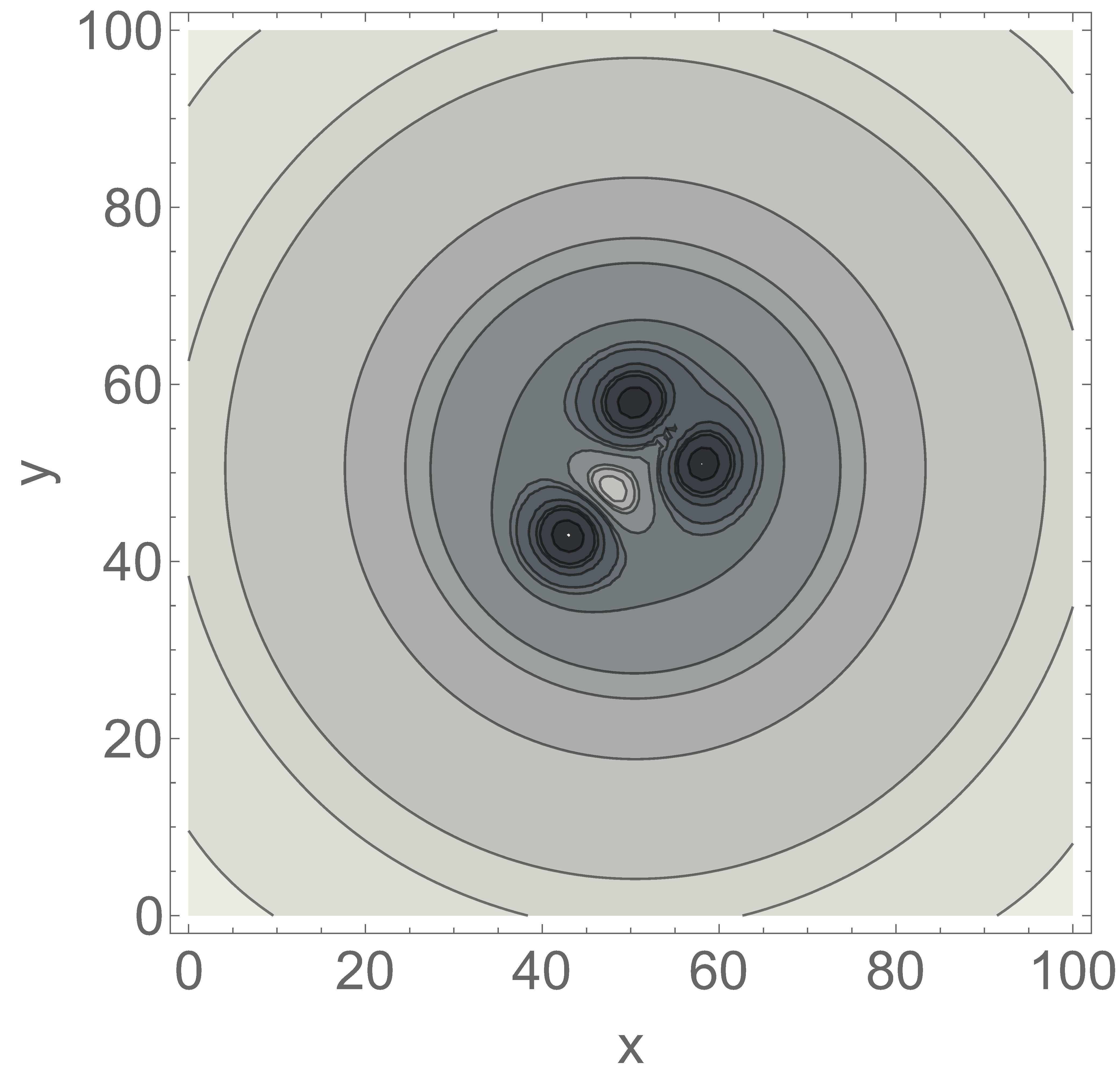}
}\hspace{8pt}
\subfigure[][]{
 \label{fig:decay_contour_c}
\includegraphics[width=0.30\hsize]{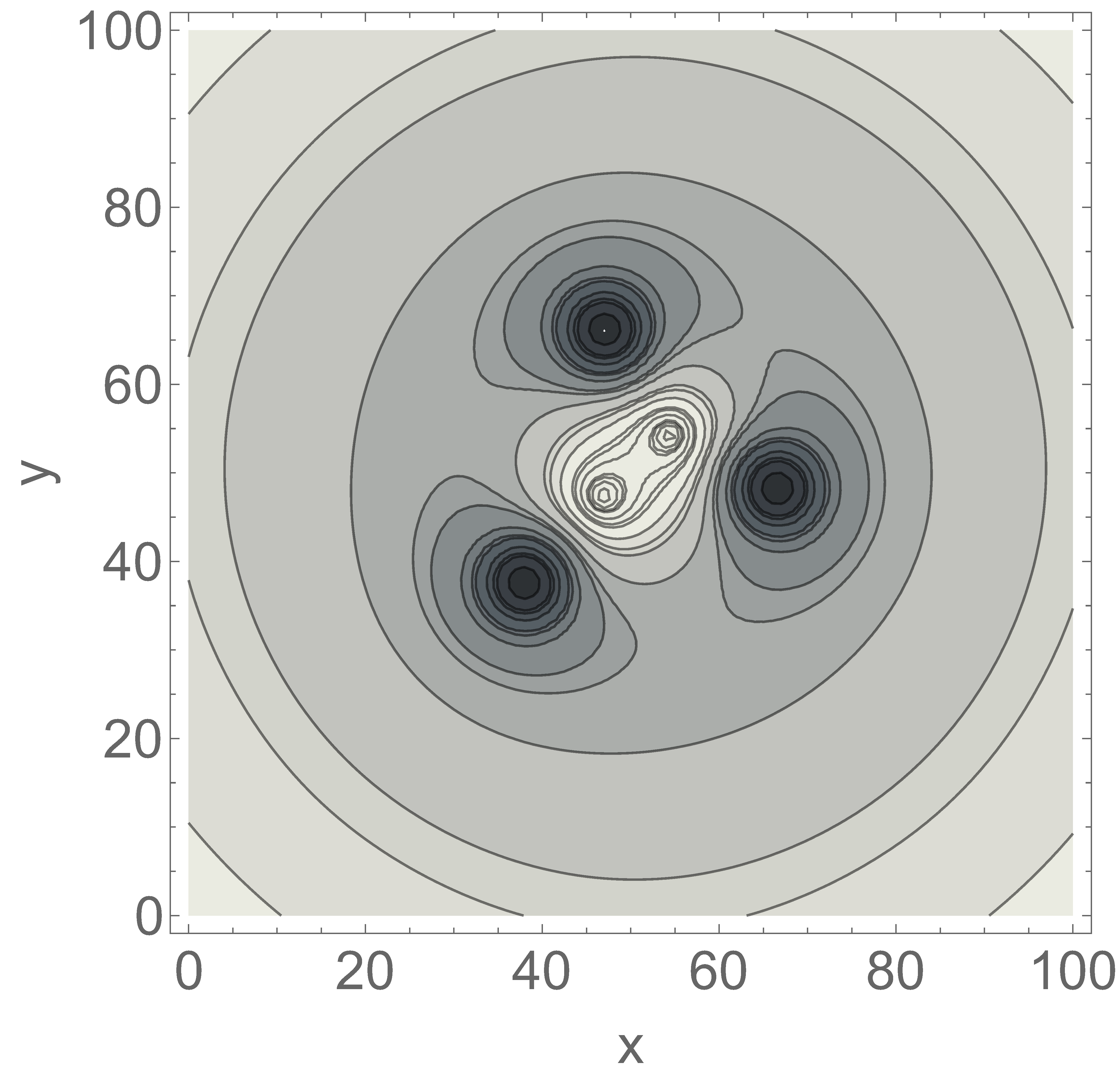}
}
\caption[]{Energy density snapshots of the time evolution of a global vortex for couplings corresponding to an unstable point in parameter space. \ref{fig:decay_contour_a} shows the initial global vortex, \ref{fig:decay_contour_b} shows the energy density just as the separation into semi-superfluid fluxtubes sets in, and \ref{fig:decay_contour_c} shows the energy density for three already well-separated fluxtubes. The center region in \ref{fig:decay_contour_b} and \ref{fig:decay_contour_c} has the lowest energy density in the system, which leads to a repulsive behavior of the semi-superfluid fluxtubes even at small distances, see discussion in the text.}
\label{fig:decay_contour}
\end{figure*}

The color-flavor-locked (CFL) phase has been suggested as a possible state of matter at very high 
density \cite{Alford:1997zt}. Because of the breaking of baryon number symmetry, the CFL phase
is a superfluid. Assuming that there are spinning neutron stars with quark matter cores in the CFL phase,
the property of superfluidity implies the existence of rotational vortices in the core. 
Even though the existence of topologically non-trivial configurations is granted by the remaining symmetry 
of the system after the CFL condensate has formed (the first homotopy group of the vacuum manifold is non-trivial \cite{Balachandran:2005ev}), global vortex solutions have been found to
be either energetically meta-stable, or energetically unstable. In the unstable case, the vortices decay into triplets of semi-superfluid strings, as discussed in \cite{Nakano:2007dq,Eto:2013hoa,Alford:2016dco}.
We study the system in the Ginzburg-Landau approach, which, apart from the gauge coupling $g$, also involves two condensate self-couplings $\lambda_1$ and $\lambda_2$, see equation (\ref{Lagrangian}) below.  
For most values of those couplings, global vortices are unstable. In particular, values relevant for neutron star core densities as derived through naive extrapolations of weak coupling results \cite{Iida:2000ha,Giannakis:2001wz}, have been found to be unstable points in the parameter space of the couplings. The relevant field configurations in a spinning neutron star with a quark matter core could therefore be semi-superfluid fluxtubes.
Let us briefly summarize the central equations for this study. 
We use an effective theory (Ginzburg-Landau), $m_u=m_d=m_s=0$, and the Lagrangian reads
\beq \label{Lagrangian}
\Lagr = \Tr \left[-\frac{1}{4} F_{ij} F^{ij} + D_{i} \Phi ^{\dagger} D^{i} \Phi   + m^2 \Phi ^{\dagger} \Phi  - \lambda_{2} (\Phi ^{\dagger} \Phi)^2 \right] - \lambda_{1}(\Tr[\Phi^{\dagger} \Phi])^2 + \dfrac{3m^{4}}{4\lambda}\ ,
\eeq
where the covariant derivative is given by $D_{i} = \partial_{i} - i g A_{i}$, $A_{i}$ is the gauge field and   $F_{ij} = \partial_{i} A_{j} - \partial_{j} A_{i} - ig  \left[ A_{i},A_{j} \right] $ the field strength tensor.
Here we introduced the coupling $\lambda$, which is given by
\beq \label{eq:lambda_def}
\lambda \equiv 3 \lambda_{1} + \lambda_{2}.
\eeq
Apart from the gauge field, there is a complex field $\Phi$ that corresponds to the CFL condensate. This matter field has a $3_c\times3_f$ matrix structure. Thus, in index notation, each matrix entry of $\Phi$ is labeled by one color index $\alpha$ and one flavor index $a$, $\phi_{\alpha a}$. 
The vacuum expectation value (vev) reads
\beq \label{eq:vev}
A_i = 0  \ , \quad  \Phi = \bar \phi \textbf{1}_{3 \times 3} \ ,
\quad \bar\phi = \sqrt{\frac{m^2}{2 \lambda}} \ .
\eeq
A global superfluid vortex can then be constructed by the following Ansatz,
\beq \label{eq:superfluid_solution}
A_i = 0  \ ,\quad
 \Phi_{\mathrm{sf}} = \bar\phi\, \beta(r) e^{i \theta} \,\textbf{1}_{3 \times 3} \ .
\eeq
Hereby, $\beta(r)$ is a solution to the differential equation
\beq \label{eq:radial_profile}
\beta^{''} + \frac{\beta^{'}}{r} - \frac{\beta}{r^{2}} - m^{2} \beta (\beta^{2} - 1) = 0 \ ,
\eeq
with the boundary conditions $\beta \rightarrow 0 \ \mathrm{as} \ r \rightarrow 0 \ $, and $\beta \rightarrow 1 \ \mathrm{as} \ r \rightarrow \infty \ $,
which provides the radial profile of a global vortex.
A typical decay of a global vortex, as well as the associated time scale (for a given value of damping in the simulation) is shown in Figure \ref{fig:decay}.
\subfiglabelskip=0pt
In Figure \ref{fig:decay_contour} we show the energy density of the global vortex and the semi-superfluid vortex after the decay. Away from the core region of the vortex, the energy density of a fluxtube is less than the energy density for the initial vortex,
\beq
\ba{rcl}
\eps_{\rm sf} &=& 3 \bar\phi^2/r^2,  \\[1ex]
\eps_{\rm ssft} &=& \third \bar\phi^2/r^2.
\ea
\label{eq:edens}
\eeq
Since three semi-superfluid fluxtubes are needed to sustain the overall winding number of one, the energy density of the triplet of semi-superfluid fluxtubes is lower by a factor of three as compared to the energy density away from the core of the global vortex. This leads to the instability, and, in general, to repulsive short range behavior, as demonstrated in Figure \ref{fig:decay_contour}.
Right after the decay (see panel \ref{fig:decay_contour_b}), the region \textit{between} the triplet of semi-superfluid fluxtubes has the lowest energy. Here we use a logarithmic grayscale to depict the energy density, where white corresponds to the lowest value the system acquired throughout the evolution, and black to the highest value. \textit{Outside} of the triplet, the arrangement still has a global winding of one, and is thus indistinguishable from the original global vortex. 
The energy far away from the global vortex is slightly higher than the center region of the triplet. Thus, the more space is covered by the core region of the semi-superfluid triplet, the lower the energy of the system. This is shown in the third panel \ref{fig:decay_contour_c}. The core area in this snapshot has increased, its bright white indicates the slightly lower energy density of this arrangement as compared to the outer region. In order to make the area of low energy density as large as possible, semi-superfluid fluxtubes thus repel. 
 
\subfiglabelskip=0pt
\section{Initial condition for simulating arrangements of semi-superfluid fluxtubes}
\label{sec-1}
In \cite{Alford:2016aiz}, we reported our progress on simulating systems with a large number of vortices in the absence of a gauge field. Here we extend the former treatment by including full gauge dynamics, which allows us to study and observe systems of large numbers of semi-superfluid fluxtubes. 
As in \cite{Alford:2016aiz}, we initialize the system with the gauge field put to zero (gauge links to unity), and choose the parameters for the couplings in such a way that the vortices are unstable. 
The initial configuration features a system of global vortices, which are set up as a superposition of $N$ vortices,
\beq \label{eq:multi_vortex_ansatz}
\Phi = \left(\prod_{i=1}^N \beta(r_i)\right)\exp\left\{i\sum_{i=1}^N n_i\theta_i\right\}\bar\phi\mathbf{1}_{3\times 3},
\eeq
where $r$ is a given point on the lattice and $i\in[1,N]$ labels the vortices, such that $r_i$ is the distance between the lattice point $r$ and the center of vortex number $i$, and $\theta_i$ is the angle between lattice point $r$ and the vortex position $i$.
In \cite{Alford:2016aiz} we faced the problem of an inappropriate boundary condition of our simulation region, which manifested itself in an attractive behavior for the global vortices. There, we treated the boundary in two different ways. \textit{Open} boundary conditions were implemented by choosing the boundary to be a copy of its neighboring points. This leads to a vanishing gradient at the boundary, which in turn lowers the total energy of the system, which finally results in an attractive behavior. Vortices in such a system would (unless there is only one vortex perfectly centered on the lattice) just leave the simulation area. This type of boundary condition is a good choice if one is interested in the short range interactions of vortices and/or fluxtubes near the center of the lattice. \textit{Fixed} boundary conditions on the other hand, freeze the field configuration of the initial setting into the boundary, such that the boundary 'remembers' the configuration that prevailed at initial time. In the case of a global vortex with winding number one at initial time, this means that the boundary 'sees' a uniform, global winding of the $\Phi$ field and a vanishing gauge field for all time. Even in the presence of more than one global vortex of winding one, a contour drawn along the boundary will just show the total number of vortices in terms of the winding number and a vanishing gauge field. Since semi-superfluid fluxtubes are local configurations, they experience the global boundary condition as (slightly) repulsive, which is good as long as the repulsion does not interfere with the fluxtube-fluxtube interactions. For global vortices, this boundary condition is appropriate if we are interested in interactions on a slightly larger scale. In this study we make use of both boundary conditions, depending on the field configuration and interaction scales of interest.
  We set up our system as follows. First, we choose a fixed number of vortices (in the result shown below we use 19). The vortices are placed at random points on the lattice, with two restrictions. First, there is a minimal separation among any two vortices placed on the lattice. In addition, a second constraint ensures that no vortex is placed too close to the boundary. This is crucial, since the global vortices repel each other and approach the boundary. In order to trigger the decay into the semi-superfluid fluxtubes, we add a small random perturbation to the matter fields at initial time (see discussion of the instability in \cite{Alford:2016dco}). Because we allow for a relatively large distance from each vortex to its nearest boundary, we do not add additional vortices \textit{beyond} the boundary, as discussed and implemented in \cite{Alford:2016aiz}.

\section{Time evolution of a multi-fluxtube arrangement}
\label{sec-2}
\begin{figure*}[ht!]
\centering
\subfigure[][]{
 \label{fig:multi_decay_a}
\includegraphics[width=0.40\hsize]{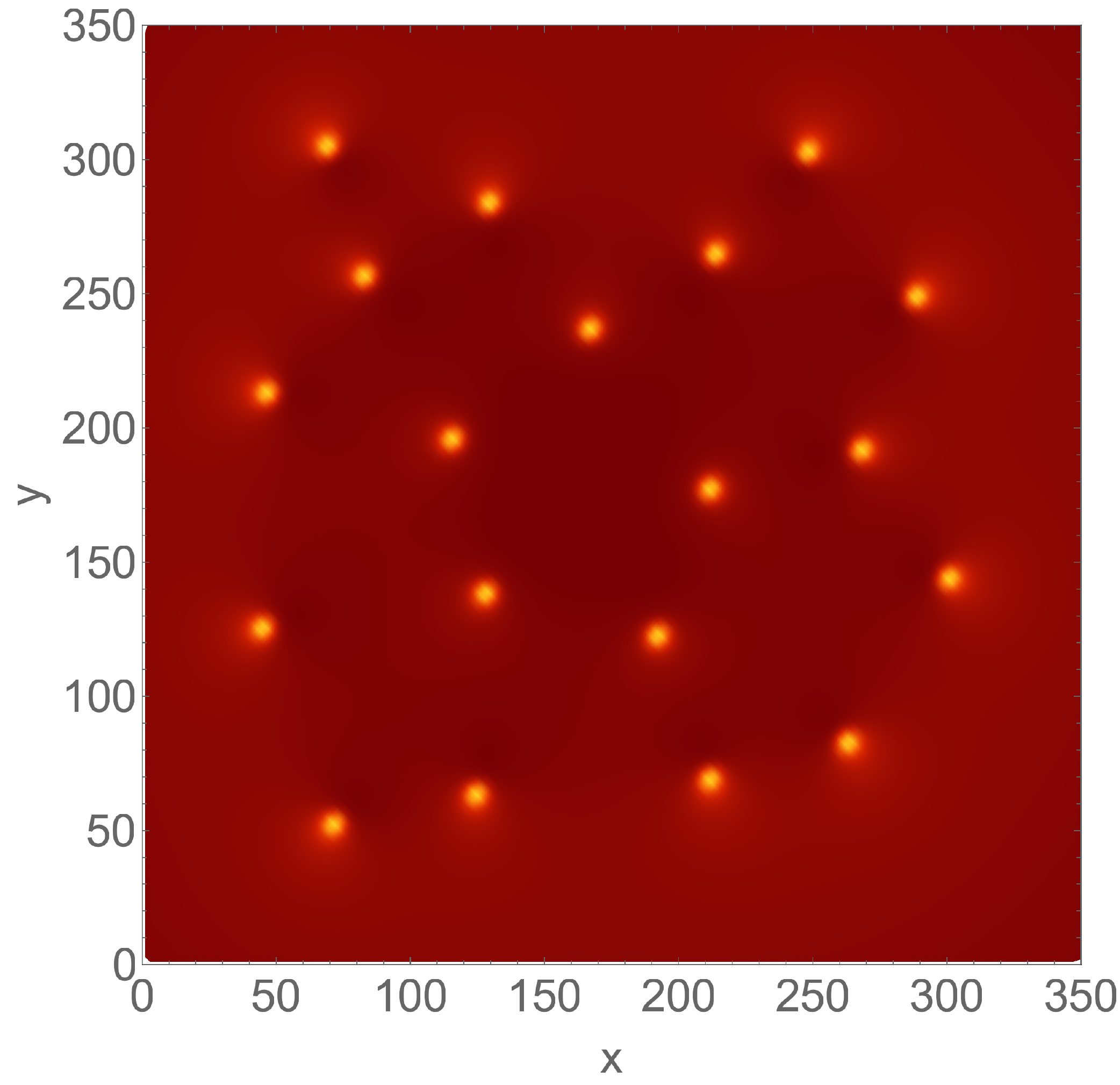}
}\hspace{8pt}
\subfigure[][]{
 \label{fig:multi_decay_b}
\includegraphics[width=0.40\hsize]{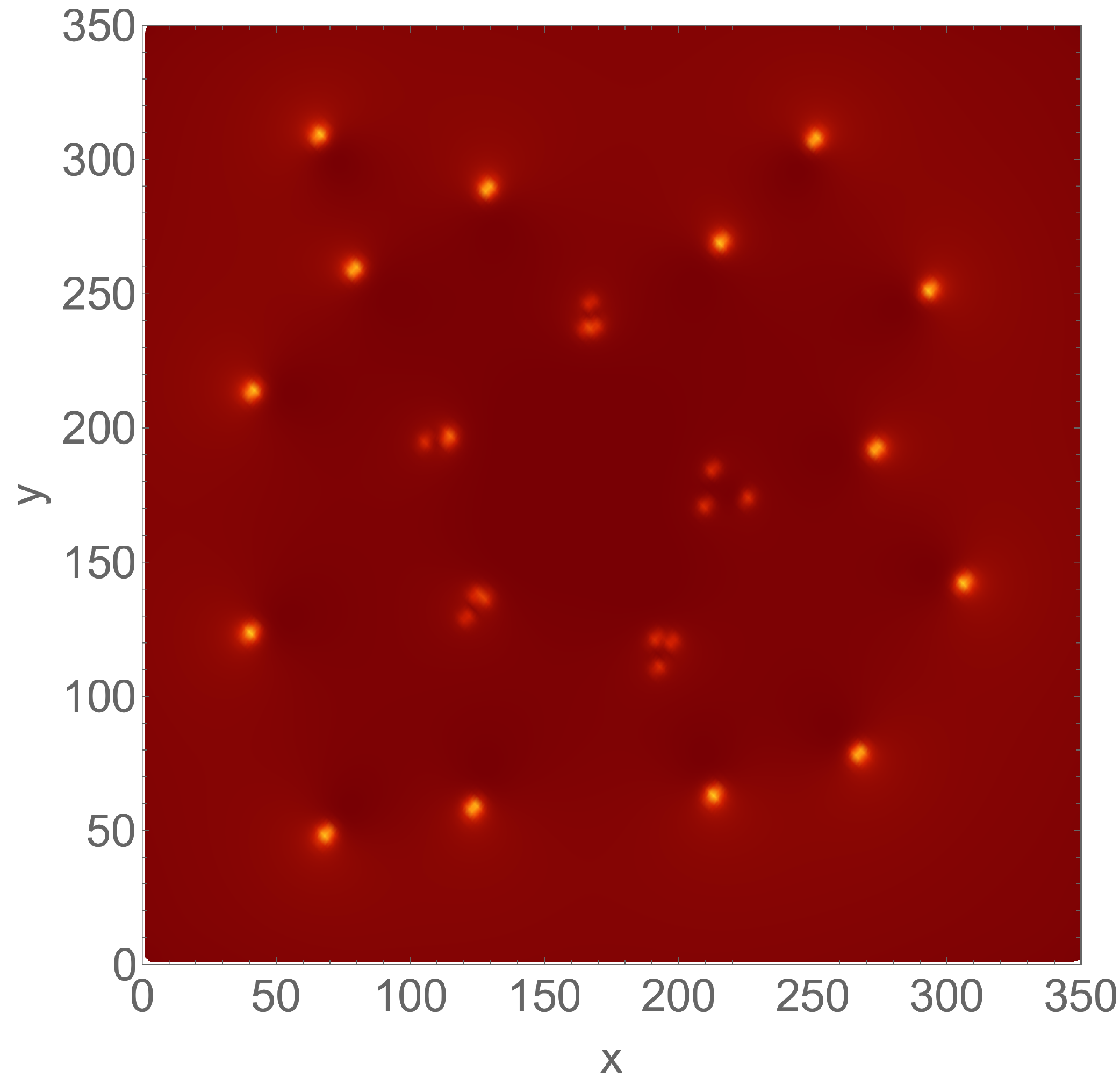}
}\\
\subfigure[][]{
 \label{fig:multi_decay_c}
\includegraphics[width=0.40\hsize]{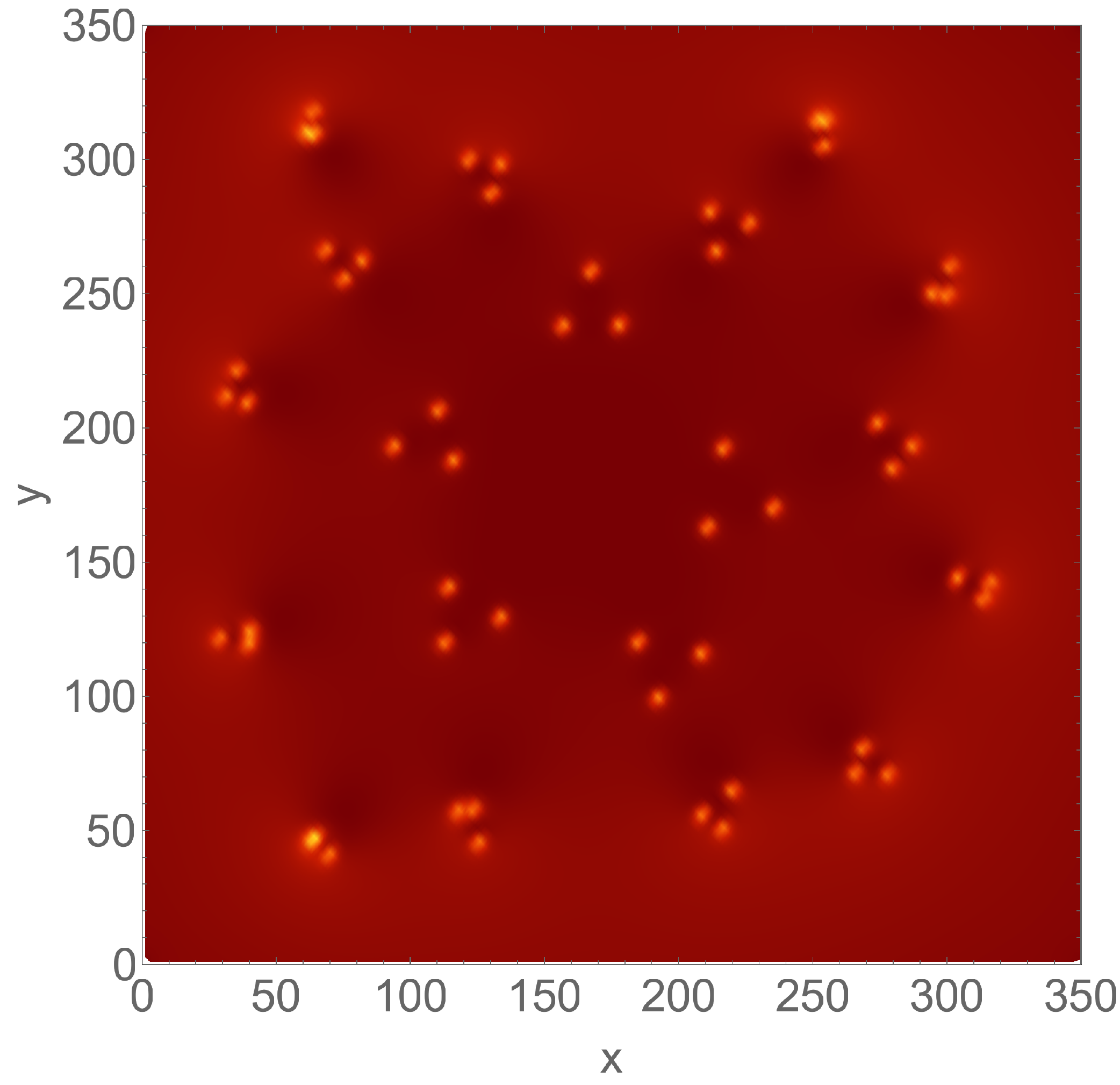}
}\hspace{8pt}
\subfigure[][]{
 \label{fig:multi_decay_d}
\includegraphics[width=0.40\hsize]{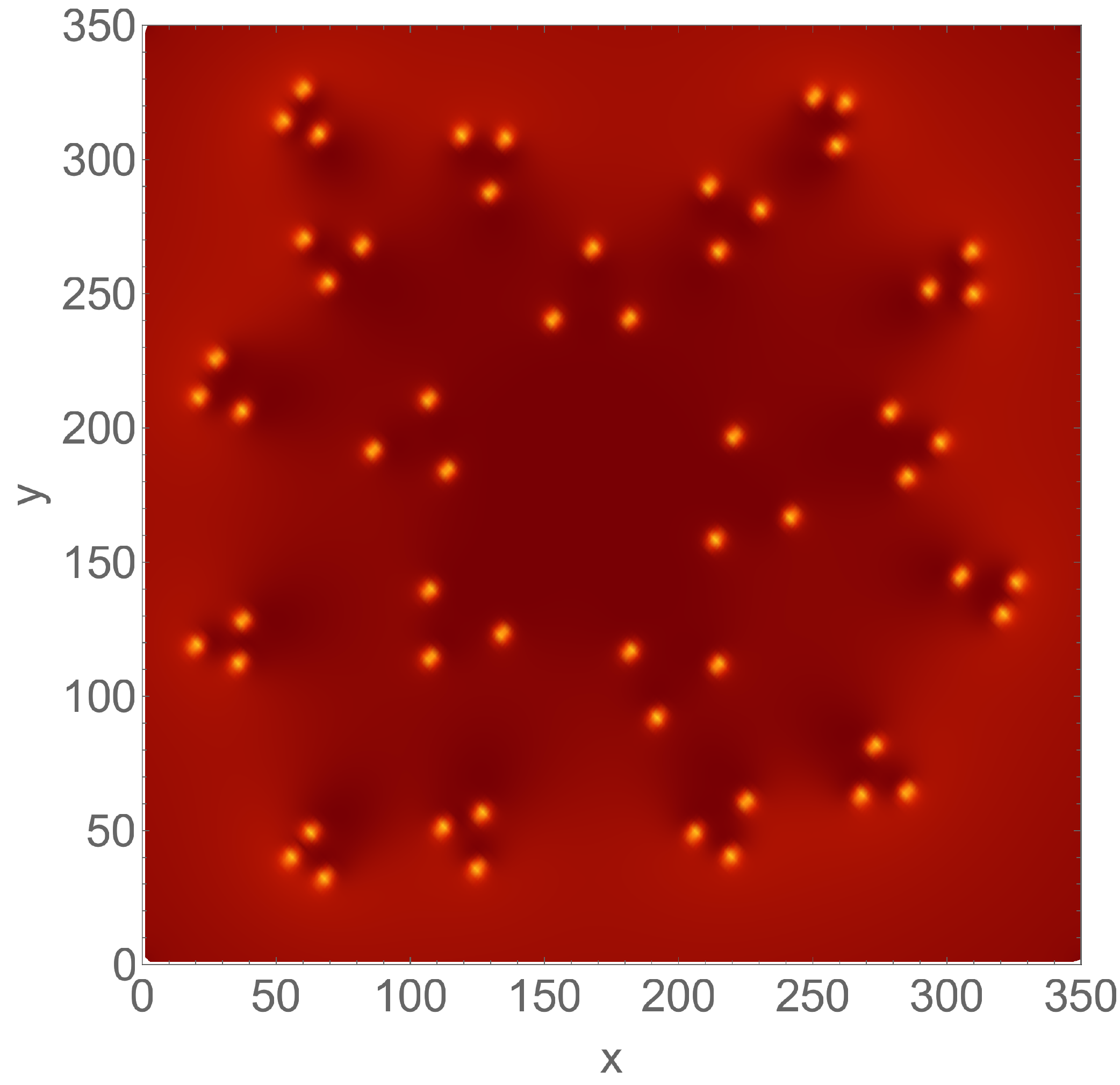}
}
\caption[]{Time evolution of the energy density of a system of 19 global vortices for an unstable point in the parameter space of the couplings. The gauge field is put to zero (gauge links unity) at initial time, but acquires non-zero values throughout the evolution. The decay into multiple triplets of semi-superfluid fluxtubes first occurs at vortices located near the center of the lattice, which could be due to the effect of the fixed boundary condition. Panel \ref{fig:multi_decay_a} shows the energy of the system at $t=0$, \ref{fig:multi_decay_b} at $t=240$, \ref{fig:multi_decay_c} at $t=480$ and \ref{fig:multi_decay_d} at $t=720$ respectively.}
\label{fig:multi_decay}
\end{figure*}

\begin{figure*}[ht!]
\centering
\subfigure[][]{
 \label{fig:multi_decay_contour_a}
\includegraphics[width=0.40\hsize]{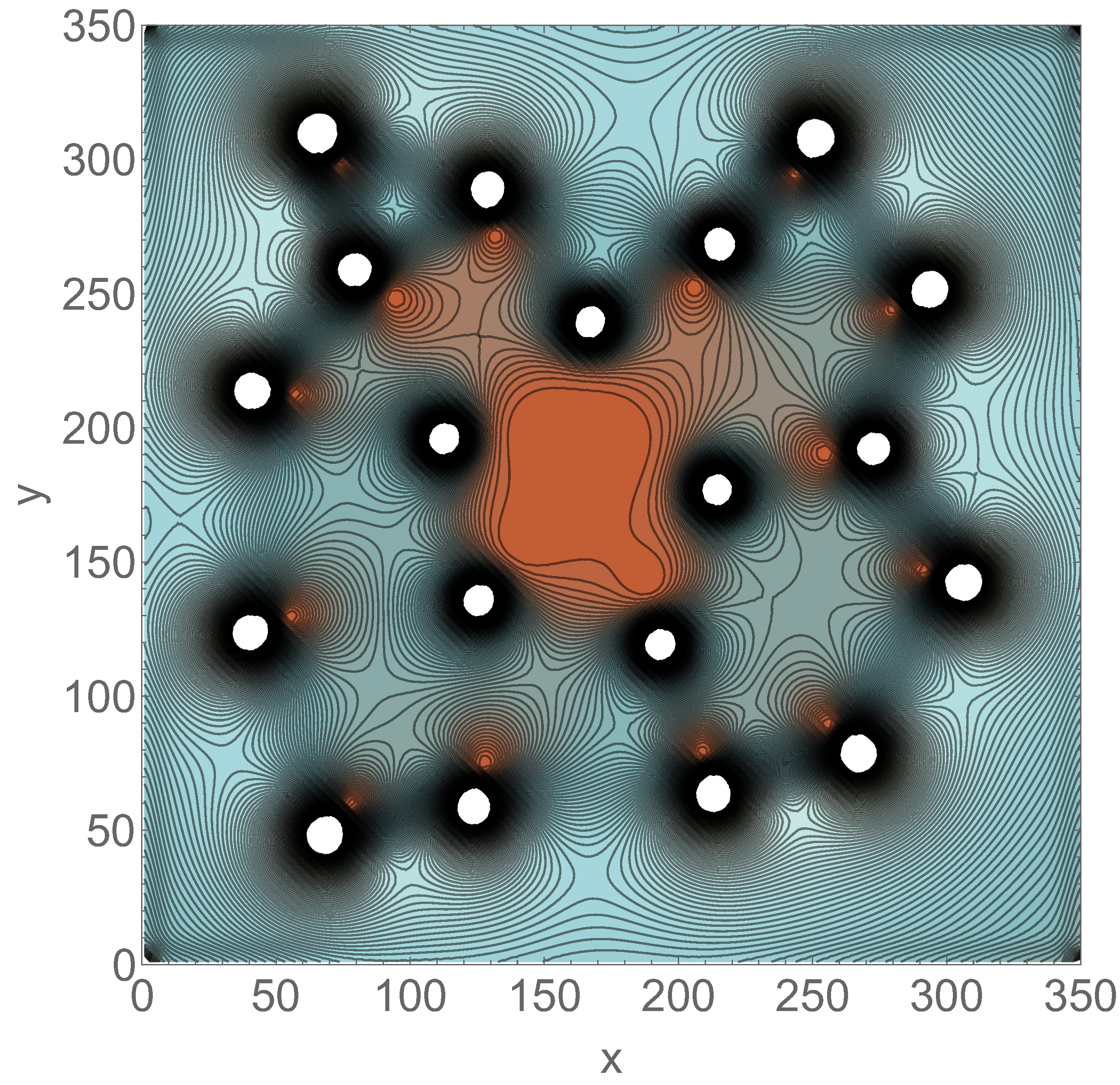}
}\hspace{8pt}
\subfigure[][]{
 \label{fig:multi_decay_contour_b}
\includegraphics[width=0.40\hsize]{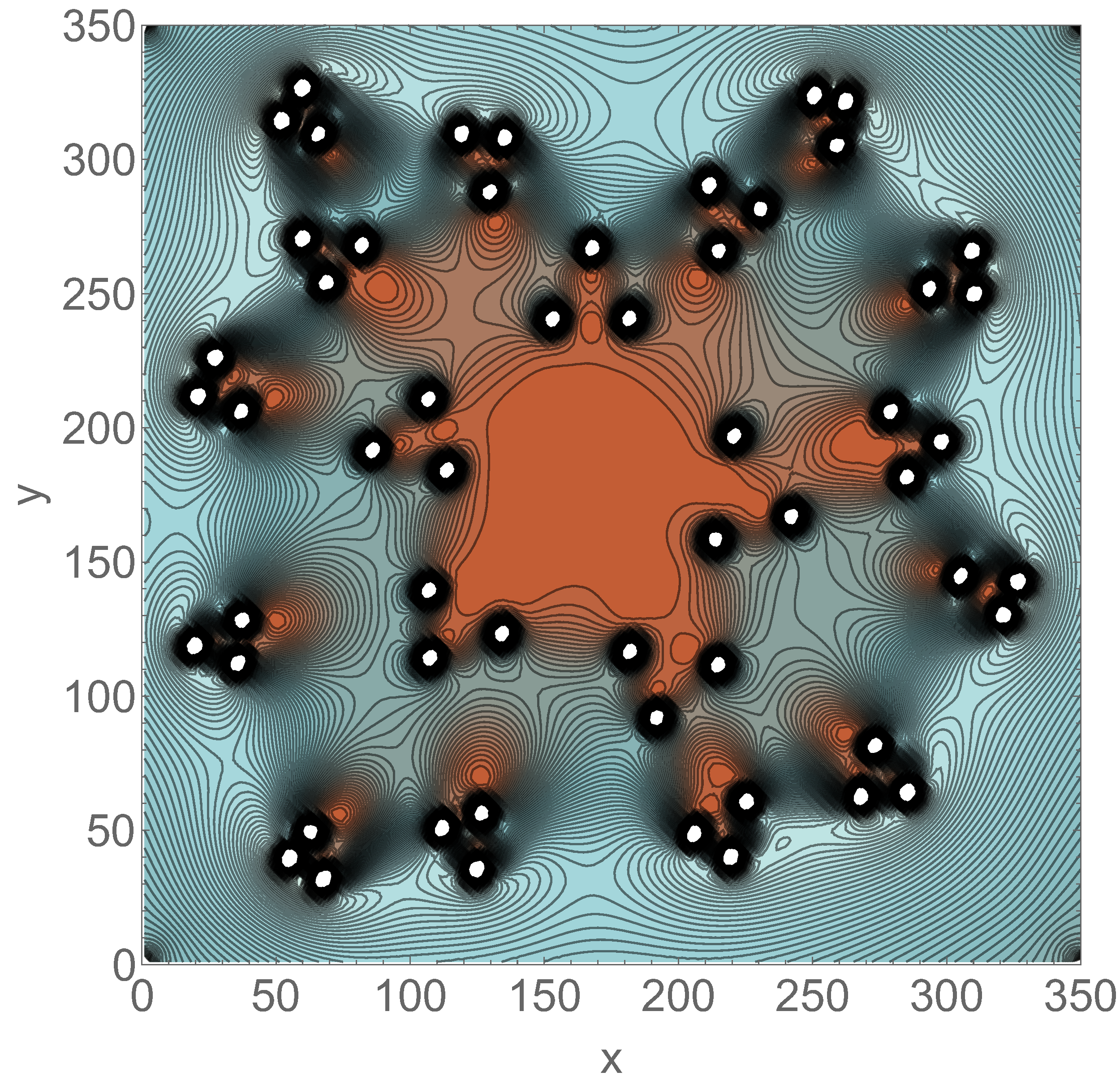}
}
\caption[]{Energy density contours of the initial and final configuration of the simulation shown in Figure \ref{fig:multi_decay} above. In panel \ref{fig:multi_decay_contour_a}, the 19 initial vortices are shown. The center region (orange in the color figure) has the lowest energy in the arrangement of the global vortices because of cancellations of winding. The vortices closer to the boundary are slightly deformed, as indicated by the small, bright (orange) speck on the side facing the center of the lattice. The final stage of this simulation, \ref{fig:multi_decay_contour_b}, shows that, as in Figure \ref{fig:decay_contour}, the semi-superfluid triplets lower the energy in their center region. They also seem to arrange themselves in a favorable manner with respect to one another. However, the huge winding number and relatively small lattice strongly affects the vortices closer to the boundary, which not only decay much later, but are more or less held in place by a steep energy gradient which arises due to the inconsistency of the boundary condition.} 
\label{fig:multi_decay_contour}
\end{figure*}

Here we discuss the time evolution of 19 global vortices, which decay into 57 semi-superfluid fluxtubes. The simulation has been performed on a $350\times 350$ lattice. Details of the initial condition are outlined in Section \ref{sec-1} above.
The energy density of the initial configuration, together with three further snapshots of the time-evolved system, are shown in Figure \ref{fig:multi_decay}. 
\subfiglabelskip=0pt
The first observation is, that vortices closer to the boundary decay later than those close to the center. This is most likely an artifact induced by the boundary, since, as discussed above, the boundary \textit{does} exert a small force on the vortices. Eventually though, all vortices decay and arrange themselves into the usual triplets of semi-superfluid fluxtubes. 
A closer look at the energy density is presented in Figure \ref{fig:multi_decay_contour}.
\subfiglabelskip=0pt
The energy density contours indicate that the lowest energy region of the initial configuration around the center of the lattice enlarges considerably throughout the evolution, since the global vortices are pushed away from the center due to their mutual repulsion (the boundary affects the vortices only very weakly, so the overall tendency of the vortices is to move outwards). 
At $t=0$, when all vortices are global, the system is perturbed through a small random mode.
Because of the initial condition being set up in a way that no initial vortex is too close to the boundary, the decay into the triplet of fluxtubes sets in before any vortex hits the boundary as it is pushed outwards. 
It is interesting to see how the semi-superfluid fluxtubes for the five vortices closest to the center arrange themselves with respect to one another. They sit at the vertices of the pentagon shaped center region, but do not seem to have separated much further from one another, as becomes evident from comparing the last two snapshots of Figure \ref{fig:multi_decay}.
Since, so far, we only performed this one simulation, we can only speculate at this point why the fluxtubes have not arranged themselves in a more homogeneous way. One possible explanation for this is, that the simulation has not been evolved to a time $t$ that is large enough to provide an equilibrated configuration. Apart from the lack of separation, it is also interesting that not one fluxtube has entered the center region of the lattice.
Here we distributed the initial vortices at random positions on the lattice.
However, another question would be, given a certain geometric arrangement of the initial vortices, how does the system arrange the semi-superfluid fluxtubes in order to lower the energy the most? 
In order to investigate this more closely, the next step will thus be to set up a, say, hexagonal lattice of global vortices and see how the decay products are oriented in such a highly symmetric structure. 
\section{Conclusions and outlook}
\label{sec-3}
In these proceedings we report the status of our extensive numerical computations that can simulate the dynamics of rotational vortices in a CFL quark matter superfluid \cite{Alford:2016dco,Alford:2016aiz}.
Currently we are in the process of extending our code in such a way that we can study large systems of multiple global vortices and semi-superfluid fluxtubes, which will provide us with a better understanding of the interaction among the fluxtubes. Ultimately, we are interested in studying pinning phenomena in the context of a superfluid quark matter core of a neutron star, which, if existent, provides a possible explanation for neutron star glitches and hence a relation between interior properties of a star with observable quantities. 
Using the experience gathered from global vortex simulations \cite{Alford:2016aiz}, we went one step further and presented first results of the time evolution of multiple global vortices, while allowing for full gauge dynamics. After perturbing the system with a small random configuration, we could observe the expected decay of the vortices into triplets of semi-superfluid fluxtubes. 
The vortices in the center region of the simulation, which were practically unaffected by the presence of the boundary, provide first insights into how the fluxtubes may arrange themselves with respect to one another in the energetically most favorable way. This simulation was, however, performed on a relatively small lattice for such a high number of vortices. Taking the gauge dynamics into account makes the simulation much more costly, since most of the simulation time is spent on updating the gauge fields. This is due to the fact that the gauge fields are represented as group valued link variables on the lattice rather than algebra valued quantities. This leads to the necessity of performing matrix exponentials at each link at every time slice in order to propagate the gauge field in time. One future task will thus be to implement the computation using graphics processing units (GPUs), which allows for a high level of parallelization. This will in turn enable us to explore systems of much larger size, which renders the existence of the boundary much less dreadful. On the less technical side, and already within reach on the small lattices we are currently running, we plan to investigate the decay pattern of symmetric arrangements of global vortices at initial time. This will provide us with insights into how lattices of semi-superfluid fluxtubes arrange themselves, which could play a role in the understanding of pinning phenomena. As a further modification of our original study \cite{Alford:2016dco}, we furthermore abandon the premise of equal masses and assume $m_u\neq m_d\neq m_s$, based on the approach presented in \cite{Eto:2009tr}. Even though we already produced some first results in this case as well, we are still investigating a peculiarity arising in the strange quark asymmetric case, whose precise nature and origin has not yet been revealed and will be the subject of a future study.

\section*{Acknowledgments}
This study has been supported by the U.S. Department of Energy, Office of Science, Office of Nuclear Physics under Award number \#DE-FG02-05ER41375. AW acknowledges support through the Austrian Science Fund (FWF) Schr\"odinger Fellowship J 3800-N27.

%
\bibliography{ssft_dynamics}

\end{document}